\documentclass[11pt,onecolumn,draftcls]{IEEEtran}
\usepackage{amsmath,amssymb,cite,multirow,subfigure}
\usepackage{graphicx}
\usepackage{url}

\newtheorem{proposition}{Proposition}

\newtheorem{lemma}{Lemma}
\newtheorem{corollary}{Corollary}

\newcommand{\df}{\stackrel{\mbox{\scriptsize def}}{=}}
\newcommand{\gf}{\mathrm{GF}}
\newcommand{\rk}{\mathrm{rk}}

\newcommand{\Ar}{A_{\mbox{\tiny{R}}}}

\newcommand{\dr}{d_{\mbox{\tiny{R}}}}
\newcommand{\ds}{d_{\mbox{\tiny{S}}}}
\newcommand{\di}{d_{\mbox{\tiny{I}}}}

\newcommand{\gr}{g_{\mbox{\tiny{R}}}}

\newcommand{\Jr}{J_{\mbox{\tiny{R}}}}
\newcommand{\Js}{J_{\mbox{\tiny{S}}}}

\newcommand{\Jc}{J_{\mbox{\tiny{C}}}}

\newcommand{\Nr}{N_{\mbox{\tiny{R}}}}
\newcommand{\Ns}{N_{\mbox{\tiny{S}}}}
\newcommand{\Ni}{N_{\mbox{\tiny{I}}}}

\renewcommand{\Pr}{P_{\mbox{\tiny{R}}}}

\newcommand{\Ps}{P_{\mbox{\tiny{S}}}}
\renewcommand{\Pi}{P_{\mbox{\tiny{I}}}}

\newcommand{\Vr}{V_{\mbox{\tiny{R}}}}

\begin{document}
\title{On the Decoder Error Probability of Rank Metric Codes and Constant-Dimension Codes}

\author{Maximilien Gadouleau,~\IEEEmembership{Member, IEEE,}
and Zhiyuan Yan,~\IEEEmembership{Senior Member, IEEE}%
\thanks{This work was supported in part by Thales Communications
Inc. and in part by a grant from the Commonwealth of Pennsylvania,
Department of Community and Economic Development, through the
Pennsylvania Infrastructure Technology Alliance (PITA).
This work was presented in part at the IEEE International Symposium on Information Theory, Seoul, South Korea, June 2009, and at the IEEE Information Theory Workshop, Taormina, Italy, October 2009.} %
\thanks{Maximilien Gadouleau is with CReSTIC, Universit\'e de Reims Champagne-Ardenne, Reims, France. Zhiyuan Yan is with the Department of Electrical and Computer
Engineering, Lehigh University, Bethlehem, PA, 18015 USA (E-mails:
maximilien.gadouleau@univ-reims.fr; yan@lehigh.edu).}}

\maketitle

\thispagestyle{empty}

\begin{abstract}
Rank metric codes and constant-dimension codes (CDCs) have been
considered for error control in random network coding. Since decoder
errors are more detrimental to system performance than decoder
failures, in this paper we investigate the decoder error probability
(DEP) of bounded distance decoders (BDDs) for rank metric codes and CDCs. For rank metric codes,
we consider a channel
motivated by network coding, where errors with the same row space are equiprobable.  Over such channels, we establish upper bounds on
the DEPs of BDDs, determine the exact DEP of BDDs for maximum rank distance (MRD)
codes, and show that MRD
codes have the greatest DEPs up to a scalar. To evaluate the DEPs of
BDDs for CDCs, we first establish some fundamental geometric
properties of the projective space. Using these geometric
properties, we then consider BDDs in both subspace and injection
metrics and derive analytical expressions of their DEPs for CDCs,
over a symmetric operator channel, as functions of their distance
distributions. Finally, we focus on CDCs obtained by lifting rank
metric codes and establish two important results: First, we derive
asymptotically tight upper bounds on the DEPs of BDDs in both
metrics; Second, we show that the DEPs for KK codes are the greatest
up to a scalar among all CDCs obtained by lifting rank metric codes.
\end{abstract}

\section{Introduction}\label{sec:intro}

Two important classes of codes for error correction in random network coding are \emph{constant-dimension codes} (CDCs) and \emph{rank metric codes}. The first class of codes are CDCs, which refer to codes defined in Grassmannians associated with the vector space over a finite field. They correct the errors and erasures which typically occur on the network by using the subspace metric \cite{koetter_it08}, while using the injection metric they correct errors on adversarial channels \cite{silva_it09}. Many bounds and constructions were proposed in the literature (see, for example, \cite{koetter_it08, etzion_it09, gadouleau_it09_cdc}), and CDCs were shown to be asymptotically optimal subspace codes \cite{gadouleau_it09_subspace}. The second class of codes are rank metric codes \cite{delsarte_jct78, gabidulin_pit0185, roth_it91}, which are related to CDCs in more than one way. First, nearly optimal CDCs can be constructed by lifting rank metric codes \cite{silva_it08}, a transformation that preserves the distance. Second, it was recently shown that CDCs are closely related to constant-rank codes \cite{gadouleau_it10_crc}. More importantly, error correction in coherent and noncoherent random network coding can be solved from a rank metric perspective \cite{silva_it08, silva_it09}. There has been a steady stream of work on rank metric codes, in particular the maximum cardinality of a code with a given
minimum rank distance was determined in \cite{delsarte_jct78, gabidulin_pit0185, roth_it91}. We refer to codes with maximum cardinality as maximum rank distance (MRD) codes, and the class of linear MRD codes proposed independently in \cite{delsarte_jct78, gabidulin_pit0185, roth_it91} as Gabidulin codes henceforth.

One critical aspect that has received little attention in the literature is the \textbf{error performance} of bounded distance decoders (BDDs) for rank metric codes and CDCs. Given a received word, a BDD either declares a failure or finds a codeword within a predetermined radius of the received word. In the latter
case, when the codeword produced by the BDD is not the sent codeword, a decoder error occurs. In many applications and especially in network coding, a decoder error is more detrimental than a decoding failure, and the decoder error probability (DEP) is a crucial parameter of BDDs for the codes used in the transmission. 


In this paper, we investigate the DEP of BDDs for rank metric codes and CDCs on channels motivated by error control in random network coding.  
The main contributions of this paper follow.
\begin{itemize}
\item We first consider a channel on matrices where all additive
errors with the same row (or column) space are equiprobable. For any
rank metric code over the equal row (or column) space channel, we
derive upper bounds on the DEP of bounded rank distance decoders.
Our results show that the DEP of BDDs for \textbf{any} rank metric
code decreases
\textbf{exponentially} with the square of its error correction capability. For MRD codes over the equal row
(or column) space channel, we derive the exact DEP of BDDs and show that MRD codes have the highest DEP up to a scalar.


\item We then consider operator channels on subspaces, more precisely symmetric operator channels, where all outputs with the same dimension and at the same subspace (or injection) distance from the input are equiprobable. Using geometric properties of balls with subspace radii, we derive the DEPs of BDDs for arbitrary CDCs over a symmetric operator channel. The analytical expressions for both metrics ultimately depend on the distance distributions of CDCs.

\item Finally, we focus on CDCs obtained by lifting rank metric codes since their distance distributions are known, and obtain two important results. First, we obtain asymptotically tight upper bounds on the DEPs of BDDs in both metrics; the upper bounds decrease exponentially with the square of the decoding radius. Second, we show that the DEPs for KK codes, which are nearly optimal CDCs \cite{koetter_it08} and can be obtained by lifting Gabidulin codes \cite{silva_it08}, are the highest up to a scalar among all CDCs obtained by lifting rank metric codes.
\end{itemize}

Our work on the error performance for CDCs is novel to the best of our knowledge. Our work on the error performance for rank metric codes differs from the work in \cite{roth_it97} in several aspects, and is a nontrivial extension of our previous work \cite{gadouleau_it08_dep}. The error performance analysis in \cite{roth_it97} was aimed at two-dimensional errors in data storage equipments and as such, assumes different channel models and considers decoder errors and decoder failures together. Our results in this paper are more general in terms of both the channel model and underlying codes than those in \cite{gadouleau_it08_dep}, and they required a more fundamental geometric approach and the use of novel techniques such as constant-rank codes \cite{gadouleau_it10_crc}. This more general investigation of decoder error performance is important, since explicit construction of optimal constant-dimension codes for arbitrary parameters is unknown and suboptimal codes are sometimes favored due to complexity or error performance. 

Our work on the error performance for rank metric codes parallels some previous works for Hamming metric codes. In \cite{mceliece_it86}, an upper bound on the DEP of a bounded Hamming distance decoder for linear MDS codes over channels where all errors with the same Hamming weight are equiprobable was derived. This work was refined in \cite{cheung_it89}, where the exact DEP for linear MDS codes was determined under the same setting. In \cite{tolhuizen_eurocode92}, the results in \cite{mceliece_it86} were extended to more general channels and to any linear code. More precisely, \cite{tolhuizen_eurocode92} introduces error-value symmetric channels, where all errors with the same support are equiprobable, thus taking bursty channels into account. Our results on the DEP 
for rank metric codes over the equal row (or column) space channel parallel those in \cite{tolhuizen_eurocode92}. We would like to emphasize that the work in this paper strongly differs from the previous work reviewed above. First, while readily available combinatorial results can be used to derive the results for Hamming metric codes (for example, the results in \cite{cheung_it89} are based on inclusion-exclusion principle), their counterparts for rank metric codes have to be established explicitly. Also, our results for rank metric codes and CDCs are based on completely different approaches from those used for Hamming metric codes. 

The rest of the paper is organized as follows. Section~\ref{sec:preliminaries} reviews some necessary background. In Section~\ref{sec:DEP_rank_metric}, we investigate the DEPs for rank metric codes. In Section~\ref{sec:DEP_CDC}, we investigate the DEPs for CDCs in general and liftings of rank metric codes in particular.

\section{Preliminaries} \label{sec:preliminaries}

\subsection{Rank Metric Codes}
\label{sec:rank_metric}
The {\em rank distance} between two matrices in $\mathrm{GF}(q)^{m \times n}$ is defined as $\dr({\bf X}, {\bf Y}) \df \rk({\bf X} - {\bf Y})$. The number of matrices at distance $u$ from a given matrix is denoted as $\Nr(u) = {n \brack u} \alpha(m,u)$\footnote{For completeness, it should be denoted as $\Nr(q, m, n, u)$. When there is no ambiguity about the underlying field and the size of matrices, we use $\Nr(u)$ for simplicity. To simplify the notation in this paper, similar simplifications have been made to other symbols.}, where $\alpha(m,u) = 0$ for $m<0$, $\alpha(m,0) = 1$, and $\alpha(m,u) = \prod_{i=0}^{u-1} (q^m-q^i)$ for $m\geq 0$ and $u \geq 1$, and ${n \brack u} = \frac{\alpha(n,u)}{\alpha(u,u)}$ is referred to as the Gaussian binomial. For all $0 \leq r \leq n$, we have
\begin{equation} \label{eq:Gaussian}
	q^{r(n-r)} \leq {n \brack r} < K_q^{-1} q^{r(n-r)},
\end{equation}
where $K_q = \prod_{j=1}^\infty (1-q^{-j})$ \cite{gadouleau_it08_dep}. The volume of a ball with rank radius $t$ in $\mathrm{GF}(q)^{m \times n}$ is denoted as $\Vr(t) = \sum_{s=0}^t \Nr(s)$. The \emph{intersection number} $\Jr(u,s,d)$, defined as the volume of the intersection of two spheres with radii $u$ and $s$ and distance $d$ between their centers, was derived in \cite{gadouleau_cl08}. In particular, the intersection numbers satisfy $\Jr(t,d-t,d) = q^{t(d-t)} {d \brack t}$ for all $0 \leq t \leq d \leq \min\{n,m\}$, and \cite{brouwer_book89}:
\begin{eqnarray}
    \label{eq:N_J}
    \Nr(d) \Jr(u,s,d) &=& \Nr(u) \Jr(d,s,u)\\
    \label{eq:sum_J}
    \sum_{u=0}^n \Jr(u,s,d) &=& \Nr(s).
\end{eqnarray}

A {\em rank metric code} can be viewed as a subset of $\mathrm{GF}(q)^{m \times n}$, where the minimum rank distance of a code is simply the minimum distance over all pairs of distinct codewords. The maximum cardinality of a rank metric code in $\mathrm{GF}(q)^{m \times n}$ with minimum rank distance $d$ is $\min\{q^{m(n-d+1)}, q^{n(m-d+1)}\}$ \cite{delsarte_jct78, gabidulin_pit0185, roth_it91}. We refer to codes with maximum cardinality as maximum rank distance (MRD) codes. The number of codewords at rank distance $r$ from a given codeword in an MRD code in $\mathrm{GF}(q)^{m \times n}$ ($n \leq m$) with minimum rank distance $d$ was determined in \cite{delsarte_jct78, gabidulin_pit0185} and is denoted as $M(d,r)$. In particular, we have $M(d,d) = {n \brack d} (q^m-1)$.

A constant-rank code is a rank metric code whose codewords have the same rank \cite{gadouleau_it10_crc}. The maximum cardinality of a constant-rank code in $\mathrm{GF}(q)^{m \times n}$ with minimum rank distance $d$ and rank $r$, denoted as $\Ar(q,m,n,d,r)$, satisfies $\Ar(q,m,n,d,r) = \Ar(q,n,m,d,r)$ and, for $n \leq m$ and $d \leq r$ \cite[Proposition 8]{gadouleau_it10_crc},
\begin{equation}\label{eq:bound_Ar}
    \Ar(q,m,n,d,r) \leq {n \brack r} \alpha(m,r-d+1).
\end{equation}

\subsection{Constant-Dimension Codes}
\label{sec:CDC}

We refer to the set of all subspaces of $\mathrm{GF}(q)^n$ with
dimension $r$ as the Grassmannian and denote it as $E_r(q,n)$. We
also refer to the projective space $\bigcup_{r=0}^n E_r(q,n)$ as $E(q,n)$. For $U, V \in E(q,n)$, their intersection $U \cap V$ is also
a subspace in $E(q,n)$, and we denote the smallest subspace
containing the union of $U$ and $V$ as $U+V$. The subspace metric
\cite{koetter_it08} and the injection metric \cite{silva_it09}, respectively defined as
\begin{eqnarray*}
	\ds(U, V) &\df& \dim(U + V) - \dim(U \cap V)\\
	\di(U,V) &\df& \frac{1}{2} \ds(U,V) + \frac{1}{2} |\dim(U) - \dim(V)| = \max\{\dim(U),\dim(V)\}- \dim(U \cap V)
\end{eqnarray*}
are both metrics over $E(q,n)$.

The number of subspaces with dimension $s$ at subspace distance $d$
from a subspace with dimension $r$ ($0\leq r \leq n$), denoted as $\Ns(r,s,d)$, is
$q^{u(d-u)} {r \brack u} {n-r \brack d-u}$ when $u =
\frac{r+d-s}{2}$ is an integer, or $0$ otherwise
\cite{gadouleau_it09_subspace}. The number of subspaces with
dimension $s$ at injection distance $d$ from a subspace with
dimension $r$ is given by $\Ni(r,s,d) = \Ns(r,s,2d-|r-s|)$ \cite{gadouleau_it09_subspace}. 
In the Grassmannian $E_r(q,n)$, the intersection of any two spheres of injection radius $u$ and $s$ with injection distance $d$ between their centers, referred to as the intersection number $\Jc(r,u,s,d)$, was determined in \cite{gadouleau_it09_subspace}.

A subset of $E_r(q,n)$ is called a constant-dimension code (CDC). Since $\ds(U,V) = \ds 2\di(U,V)$ if $U$ and $V$ have the same dimension, the minimum subspace distance of a CDC is equal to twice its minimum injection distance.
CDCs are related to rank metric codes through constant-rank codes or through the
lifting operation \cite{silva_it08}, described below. The lifting of ${\bf C} \in
\mathrm{GF}(q)^{r \times (n-r)}$ is defined as $I({\bf C}) = R({\bf
I}_r | {\bf C}) \in E_r(q,n)$, where ${\bf I}_r$ is the $r \times r$
identity matrix and $R$ denotes the row space of a matrix. For all ${\bf C}, {\bf D} \in \mathrm{GF}(q)^{r
\times (n-r)}$, we have $\ds(I({\bf C}), I({\bf D})) = 2\di(I({\bf C}), I({\bf D})) = 2\dr({\bf C},
{\bf D})$ \cite{silva_it08}. Therefore, the injection distance
distribution of the lifting is equal to the rank distance
distribution of the original code. Liftings of MRD codes were
introduced in \cite{silva_it08}, and we refer to them as KK codes.

\section{Decoder error probability of rank metric codes}
\label{sec:DEP_rank_metric}

In this section, we investigate the DEP of bounded rank distance decoders for rank metric codes. All distances are rank distances in this section.

We assume the following scenario, where an adversary injects \textbf{linearly independent} packets maliciously on the network, using some knowledge about the transmitted packets or the protocol used. Hence the adversary may choose and inject linearly independent packets so as to corrupt the transmitted message more effectively than others (see the example in \cite{silva_it09}). These packets undergo linear combinations through the network, and result into an additive error at the receiver whose rank depends on the number of packets injected. Due to the vector-space preserving property of linear network coding, the row space spanned by these packets remains unchanged through the linear combinations operated at intermediate nodes. Hence, the additive error at the receiver can take any value provided its row space is fixed. This leads to our model of additive errors, where all errors with the same row space are equiprobable. Also, because the rank is preserved by transposition, we also consider channels where errors with the same column space are equiprobable. A channel on $\mathrm{GF}(q)^{m \times n}$ is hence said to be \textit{equal row (column) space} if errors are additive and the errors with the same row (column) space are equiprobable.

Although equal row and column space channels were proposed to model error correction in random linear network coding as described in Section I, it is remarkable that these channels may also be used to model other applications of rank metric codes. Rank metric codes can be used for the correction of two-dimensional errors \cite{gabidulin_pit0285, roth_it91} (i.e., errors confined to a certain number of rows and columns) in storage equipments. Hence, our model encompasses the case of two-dimensional errors, where some rows or columns are more likely to be in error than others.

Note that using a code $\mathcal{C}$ over an equal row space channel on $\mathrm{GF}(q)^{m \times n}$ is equivalent to using its transpose code over a column space channel on $\mathrm{GF}(q)^{n \times m}$. Since the transpose operation preserves the distance, we only study equal row space channels. In this case, the DEP depends on the sent codeword ${\bf C}$, the row space $U \in E_u(q,n)$ of the error, and the error correction capability $t$ of the code. We hence derive a bound on $\Pr({\bf C}, U, t)$ for all codes, and we also obtain the exact value of the DEP for linear MRD codes when $n\leq m$. Since the distance distribution of an MRD code does not depend on the codeword, let us denote the DEP of a BDD for an MRD code in $\mathrm{GF}(q)^{m \times n}$ with minimum distance $d$ as $P_{\tiny \mbox{R,MRD}}(U, t)$.

Let $\Pr(\mathcal{C},t)$ the DEP of a bounded distance decoder with radius $t$ for the code $\mathcal{C}$. We have
$$
	\Pr(\mathcal{C},t) = \sum_{{\bf C} \in \mathcal{C}} \sum_{U \in E(q,n)} \Pr({\bf C},U,t) \mathrm{P}\{{\bf C},U\} \leq \max_{U, {\bf C}} \Pr({\bf C},U,t),
$$
where $\mathrm{P}\{{\bf C},U\}$ is the probability that the transmitted codeword is ${\bf C}$ and that the channel error has row space $U$. Therefore, although the probability $\Pr({\bf C},U,t)$ is conditional, it can be used as an upper bound on the unconditional DEP. The upper bounds on $\Pr({\bf C},U,t)$ that we shall derive can hence be applied for the overall DEP of any rank metric code.

For any ${\bf R} \in \mathrm{GF}(q)^{m \times n}$, we denote the number of matrices with row space $W$ and at rank distance $s$ from ${\bf R}$ as $\gr(W, s, {\bf R})$. We prove below that $\gr(W, s, {\bf R})$ depends on ${\bf R}$ only through its row space.

\begin{lemma}\label{lemma:g}
For all ${\bf R}, {\bf S} \in \mathrm{GF}(q)^{m \times n}$ with the
same row space $U$, $\gr(W, s, {\bf R}) = \gr(W,s, {\bf S})$.
\end{lemma}

\begin{proof}
Suppose ${\bf X} \in \mathrm{GF}(q)^{m \times n}$ has row space $W$
and satisfies $\rk({\bf X} - {\bf R}) = s$. Expressing ${\bf S} =
{\bf A} {\bf R}$ where ${\bf A} \in \mathrm{GF}(q)^{m \times m}$ has
full rank, the matrix ${\bf Y} = {\bf A} {\bf X}$ has row space $W$
and satisfies $\rk({\bf Y} - {\bf S}) = \rk({\bf X} - {\bf R}) = s$.
Thus $\gr(W, s, {\bf S}) \geq \gr(W, s, {\bf R})$. Using ${\bf R} = {\bf
A}^{-1} {\bf S}$, we show that $\gr(W, s, {\bf S}) \leq \gr(W, s, {\bf
R})$; hence $\gr(W,s,{\bf R}) = \gr(W,s,{\bf S})$.
\end{proof}
Since $\gr(W,s,{\bf R})$ depends on ${\bf R}$ only through its row space $U$, we denote it as $\gr(W,s,U)$ henceforth. The DEP is derived in terms of the distribution of codewords according to their row space $A_W({\bf C}) \df |\{ {\bf D} \in \mathcal{C}: R({\bf D} - {\bf C}) = W \}|$ for all $W \in E(q,n)$ and the $\gr(W,s,U)$ constants in Proposition \ref{prop:PER_row} below.

\begin{proposition}\label{prop:PER_row}
Assuming a codeword ${\bf C} \in \mathcal{C}$, a code in
$\mathrm{GF}(q)^{m \times n}$ with minimum distance $d$, is sent
over an equal row space channel and the channel error has row space
$U \in E_u(q,n)$, the DEP of a BDD  with decoding radius $t = \lfloor \frac{d-1}{2}
\rfloor$ is $0$ when $u < d-t$, and when
$u \geq d-t$ satisfies
\begin{eqnarray} \label{eq:PER_row_exact}
    \Pr({\bf C}, U, t) &=& \frac{1}{\alpha(m,u)} \sum_{w=d}^n \sum_{W \in E_w(q,n)} A_W({\bf C}) \sum_{s=0}^t \gr(U,s,W)\\
	\label{eq:PER_row_bound}
    &\leq& \frac{1}{\Nr(u)} \sum_{w=d}^n {n \brack w}
    \alpha(m,w-d+1) \sum_{s=0}^t \Jr(u,s,w)\\
    \label{eq:PER_row_asymptotic}
    &<& \left\{ \begin{array}{ll} K_q^{-2} q^{-t(m-n+t)} & \mbox{when } d=2t+1,\\
    K_q^{-2} q^{-t(m-n+t)-m} & \mbox{when } d=2t+2. \end{array} \right.
\end{eqnarray}
Furthermore, if $n\leq m$ and $\mathcal{C}$ is a linear MRD code, then the DEP is given
by
\begin{equation}\label{eq:PER_row_MRD}
    P_{\tiny \mbox{R,MRD}}(U, t) = \frac{1}{\Nr(u)} \sum_{w=d}^n M(d,w)
    \sum_{s=0}^t \Jr(u,s,w).
\end{equation}
\end{proposition}

The proof of Proposition~\ref{prop:PER_row} is given in Appendix~\ref{app:prop:PER_row}. We remark that the upper bound in (\ref{eq:PER_row_bound}) actually depends on $u$ only, not $U$, while the bound in (\ref{eq:PER_row_asymptotic}) is very general: it does not depend on the transmitted codeword ${\bf C}$ and  the rank of error $u$, and applies to \textbf{any rank metric code}. In fact, applying (\ref{eq:PER_row_asymptotic}) to linear MRD codes leads to \cite[Proposition 6]{gadouleau_it08_dep}. 


We now show that in all nontrivial cases MRD codes have the \textbf{greatest} DEP among all rank metric codes up to a scalar.

\begin{corollary} \label{cor:PER_symmetric_MRD}
Let $\mathcal{C}$ be any rank metric code in $\mathrm{GF}(q)^{m \times n}$ ($n \leq m$) with minimum distance $d$ and let ${\bf C} \in \mathcal{C}$. Then if $q>2$, $n<m$, or $d \neq m-1$, $\Pr({\bf C}, U, t) < H_q P_{\tiny \mbox{R,MRD}}(U, t)$, where $H_2 = 3.5$ and $H_q = \frac{q-1}{q-2}$ for $q > 2$.
\end{corollary}

\begin{proof}
By \cite[Proposition 15]{gadouleau_it10_crc}, we have $H_q M(d,r) > \Ar(d,r)$ for $n\geq r \geq d$, provided that $q>2$, $n<m$, or $d \neq m-1$. Hence $H_q P_{\tiny \mbox{R,MRD}}(U, t) > \frac{1}{\Nr(u)} \sum_{w=d}^n \Ar(d,w) \sum_{s=0}^t \Jr(u,s,w) \geq \Pr({\bf C}, U, t)$.
\end{proof}

We remark that Corollary~\ref{cor:PER_symmetric_MRD} does not hold only if $q=2$, $n=m$, and $d=m-1=n-1$, which is a trivial case. Corollary~\ref{cor:PER_symmetric_MRD} indicates that there is a tradeoff between decoder capability (or radius) and decoder error probability. Given $m$, $n$, and the dimension, MRD codes achieve the greatest minimum rank distance and hence have the greatest decoder capability. However, they also have the \textbf{greatest}
DEP among all rank metric codes up to a scalar.

We would like to emphasize our choice to consider the DEP of any rank metric code, and not only MRD codes for which an efficient BDD is known. In our scenario, the DEP corresponds to a security criterion: it is the probability that the adversary manages to thwart the communication. A low DEP is hence equivalent to a high protection against the attacks by the adversary. Since MRD codes have the highest DEP up to a scalar, they may not provide a strong enough level of security. It is hence reasonable in applications to consider another class of suboptimal codes in order to increase the level of security. Although decoding algorithms have been proposed for MRD codes only, we show below that suboptimal codes could also be used with a low-complexity decoder. Such codes can be easily designed by taking Gabidulin codes over a subspace. Also, in his seminal paper, Gabidulin introduced a whole class of codes, referred to as $q$-cyclic codes, which are analogues of cyclic codes. Amongst these codes, one can find the analogues of BCH codes and in particular, the analogues of Reed-Solomon codes which are Gabidulin codes. Since many algorithms analogous to the ones for Reed-Solomon codes have been proposed for Gabidulin codes, the analogues of BCH codes are likely to also have low-complexity decoders. Using such suboptimal codes may yield a much lower DEP, and hence a higher level of security.

In the particular case of a channel where all errors with the same rank are equiprobable (referred to as rank symmetric in \cite{gadouleau_it08_dep}), the DEP only depends on the rank $u$ of the error. It can be easily shown from (\ref{eq:PER_row_exact}) that it is related to the distance distribution $A_w({\bf C}) = |\{{\bf D} \in \mathcal{C} : \rk({\bf D} - {\bf C}) = w\}| = \sum_{W \in E_w(q,n)} A_W({\bf C})$ by
\begin{equation}
    \label{eq:PER_symmetric_exact}
    \Pr({\bf C}, u, t) = \frac{1}{\Nr(u)} \sum_{w=d}^n A_w({\bf C}) \sum_{s=0}^t \Jr(u,s,w).
\end{equation}
The result in (\ref{eq:PER_symmetric_exact}) is interesting, as the distance distribution of a code has been widely studied, and the distance distribution is usually known for a given code, while the distribution of codewords according to their row space appearing in (\ref{eq:PER_row_exact}) has received much less attention.

In many applications, the probability that the received matrix is at distance $u$ from the sent codeword decreases rapidly with $u$, and hence the overall DEP can be approximated by $\Pr({\bf C}, d-t,t)$. We note that $d-t$ is the \textbf{smallest} value for $u$ that will lead to a decoder error. Furthermore, when $u=d-t$, the channel output can only be decoded to codewords that are at distance $d$ from the sent codeword.

\begin{proposition} \label{prop:PER_d-t}
Assume a codeword ${\bf C} \in \mathcal{C}$, a code in $\mathrm{GF}(q)^{m \times n}$ with minimum distance $d$, is sent over a rank symmetric channel and the channel output is at distance $u=d-t$ from ${\bf C}$, the DEP of a BDD  with decoding radius $t = \lfloor \frac{d-1}{2} \rfloor$ is given by
\begin{equation} \label{eq:PER_d-t}
    \Pr({\bf C}, d-t, t) = q^{t(d-t)} \frac{{d \brack t}}{{n \brack d-t} \alpha(m,d-t)} A_d({\bf C}).
\end{equation}
In particular, the DEP for an MRD code satisfies
\begin{equation} \label{eq:PER_d-t_MRD}
    P_{\tiny \mbox{R,MRD}}(d-t, t) > \left\{ \begin{array}{ll} K_q q^{-t(m-n+t)} & \mbox{when } d=2t+1,\\
    K_q q^{-t(m-n+t)-m} & \mbox{when } d=2t+2. \end{array} \right.
\end{equation}
\end{proposition}

\begin{proof}
Since the channel output can be decoded to only codewords at distance $d$ from the sent codeword ${\bf C}$, (\ref{eq:PER_row_bound}) reduces to $\Pr({\bf C}, d-t, t) = \frac{\Jr(d-t,t,d)}{\Nr(d-t)} A_d({\bf C})$, which gives (\ref{eq:PER_d-t}). For an MRD code, (\ref{eq:PER_row_MRD}) and the bounds on the Gaussian binomial in (\ref{eq:Gaussian}) yield (\ref{eq:PER_d-t_MRD}).
\end{proof}

When $u=d-t$, Proposition~\ref{prop:PER_d-t} above not only provides the DEP for any code,  but also shows that the upper bound on the DEP for MRD codes in (\ref{eq:PER_row_asymptotic}) is \textbf{tight} up to a scalar since the upper bound in (\ref{eq:PER_row_asymptotic}) and the approximation in (\ref{eq:PER_d-t_MRD}) differ by only a scalar which tends to $1$ wen $q$ increases.

%
%
%

\section{DEP for CDCs over a symmetric operator channel}
\label{sec:DEP_CDC}

\subsection{Further Properties of Balls with Subspace Radii} \label{sec:balls}

Properties of balls in $E(q,n)$ with subspace or injection radii
were investigated in \cite{gadouleau_it09_subspace}. In this
section, we determine further properties of such balls, which will
be instrumental in our analysis of DEP for CDCs.

We study the properties of balls with subspace radii only, as properties of balls with injection radii will not be useful to our derivation of the DEP. 
Proposition \ref{prop:Js} below shows that the intersection of two spheres in the projective space only depends on the radii of the spheres, the distance between between the centers, and the dimensions of the centers.

\begin{proposition} \label{prop:Js}
For all $A \in E_a(q,n), B \in E_b(q,n)$ with $\ds(A, B) = w$, the number of subspaces $C \in E_c(q,n)$ such that $\ds(A,C) = u$ and $\ds(B,C) = s$ only depends on $u$, $s$, $w$, $a$, $b$, and $c$. It is hence denoted as $\Js(u,s,w;a,b,c)$.
\end{proposition}

\begin{proof}
Let $t = \dim(A \cap B) = \frac{1}{2}(a+b-w)$ and ${\bf v}_0, {\bf v}_1, \ldots, {\bf v}_{a+b-t-1}$ be linearly independent vectors such that ${\bf v}_0, {\bf v}_1, \ldots, {\bf v}_{t-1} \in A \cap B$, ${\bf v}_t, {\bf v}_{t+1}, \ldots, {\bf v}_{a-1} \in A$, and ${\bf v}_a, {\bf v}_{t+1}, \ldots, {\bf v}_{a+b-t-1} \in B$. The matrix ${\bf V} \in \gf(q)^{(a+b-t) \times n}$ whose rows are given by the ${\bf v}_i$s has full rank, therefore there exists a nonsingular matrix ${\bf X} \in \gf(q)^{n \times n}$ such that ${\bf V} {\bf X} = {\bf I}_{b+a-t}$. Then $A{\bf X} = \{{\bf u} {\bf X} : {\bf u} \in A\}$ is the span of the unit vectors ${\bf e}_0, {\bf e}_1, \ldots, {\bf e}_{a-1}$, denoted as $I_a$ and $B{\bf X}$ is the span of ${\bf e}_0, {\bf e}_1, \ldots, {\bf e}_{t-1}, {\bf e}_a, \ldots, {\bf e}_{a+b-t-1}$, denoted as $J_b$. Therefore, $\ds(I_a, J_b) = \ds(A,B) = w$ and
\begin{eqnarray}
	\nonumber
	&&|\{ C \in E_c(q,n) | \ds(A,C) = u, \ds(B,C) = s \}|\\
	\nonumber
	&&= |\{ C{\bf X} \in E_c(q,n) | \ds(A{\bf X},C{\bf X}) = u, \ds(B{\bf X},C{\bf X}) = s \}|\\
	\nonumber
	&&= |\{ Y \in E_c(q,n) | \ds(I_a,Y) = u, \ds(J_b,Y) = s \}|,
\end{eqnarray}
which is a function of $u$, $s$, $w$, $a$, $b$, and $c$.
\end{proof}

The intersection of spheres with subspace radii is related to their volume in Corollary~\ref{cor:Js} below.

\begin{corollary} \label{cor:Js}
For all parameter values,
\begin{eqnarray}
    \label{eq:scaling_Js}
    \Ns(a,b,w) \Js(u,s,w;a,b,c) &=& \Ns(a,c,u) \Js(w,s,u;a,c,b),\\
    \label{eq:sum_Js}
    \sum_{u=0}^n \Js(u,s,w;a,b,c) &=& \Ns(b,c,s).
\end{eqnarray}
\end{corollary}

\begin{proof}
Let $A \in E_a(q,n)$. By counting the number of pairs of subspaces $(B, C)$ such that $\dim(B) = b$, $\dim(C) = c$, $\ds(A, B) = w$, $\ds(A, C) = u$, and $\ds(B,C) = s$ in two different ways, we obtain (\ref{eq:scaling_Js}). Next, let $B \in E_b(q,n)$ be fixed; we denote $\ds(A,B)$ by $w$. As $\sum_{u=0}^n J_S(u,s,w;a,b,c)$ is the number of subspaces $C \in E_c(q,n)$ such that $\ds(B,C) = s$, this sum equals $\Ns(b,c,s)$.
\end{proof}

Although the value of $\Js(u,s,w;a,b,c)$ is unknown in general, we
remark that for $a = b = c$, $\Js(2u,2s,2w;a,a,a) = \Jc(a,u,s,w)$,
the intersection number for $E_a(q,n)$. We
also determine its value when $w = u+s$ below.

\begin{proposition} \label{prop:Js_triangle}
$\Js(u,s,u+s;a,b,c)=0$ when $u > \min\{a+c,a+2b-c\}$, $s >
\min\{b+c, 2a+b-c\}$, or $u+s > \min\{a+b, n\}$. Otherwise, we have
\begin{equation}
    \Js(u,s,u+s;a,b,c) = {\frac{a-b+u+s}{2} \brack
    \frac{c-b+s}{2}} {\frac{b-a+u+s}{2} \brack \frac{c-a+u}{2}}.
\end{equation}
\end{proposition}

The proof of Proposition \ref{prop:Js_triangle} is given in Appendix \ref{app:prop:Js_triangle}.

\subsection{DEP of Bounded Subspace Distance Decoders for CDCs}
\label{sec:PES}

We study the DEP of a bounded distance decoder for a CDC
in $E_r(q,n)$ over a {\em symmetric operator channel},
defined below. An operator channel is a channel where the inputs and
outputs are subspaces in $E(q,n)$. As assumed in
\cite{koetter_it08}, the channel may erase some dimensions of the
transmitted subspace as well as inject some erroneous dimensions. We
refer to these as erasures and errors, respectively. We say an
operator channel is \emph{symmetric} when all
outputs corresponding to $\epsilon$ errors and $\rho$ erasures are
equiprobable. If the input
has dimension $r$ and $\epsilon$ errors and $\rho$ erasures occur,
the output has dimension $v = r+\epsilon - \rho$ and is at subspace
distance $u = \epsilon + \rho$ and at injection distance $\mu = \max\{\epsilon,\rho\}$ from the input.  Therefore, an operator channel is symmetric if and only if all outputs with the same dimension and at the same subspace (or injection) distance from the input are equiprobable.

Since the distance properties of a code $\mathcal{C}$ in $E_r(q,n)$
are equal to those of the code in $E_{n-r}(q,n)$ consisting of the dual subspace of each codeword in $\mathcal{C}$, we assume $2r\leq n$ as in \cite{koetter_it08}.

We first study the DEP of a bounded subspace distance decoder.  We remark that
$u+v-r=2\epsilon$ should be an even integer, otherwise there are no
subspaces with dimension $v$ and at distance $u$ from the
transmitted subspace, that is, $\Ns(r,v,u) = 0$.

For a CDC $\mathcal{C} \subseteq E_r(q,n)$ with minimum subspace
distance $2d$, suppose a codeword $C$ is transmitted over a
symmetric operator channel. The output of a BDD with
decoding radius $d-1$ and its DEP depend on both $\epsilon$ and
$\rho$, or equivalently on both $u$ and $v$. First, if $u\leq d-1$,
that is, the channel output is at subspace distance up to $d-1$ from
the sent codeword $C$, the BDD will produce the sent codeword $C$.
When $u = d$, the channel output is beyond the decoding radius of
any codeword, and hence the BDD will produce a failure. When $u\geq
d+1$ and $|v-r| > d-1$, since the distance between the received subspace and the code is no less than $|v-r|$, the channel output is beyond the decoding radius of any codeword and the BDD will produce a failure. In all the cases above, the DEP is zero.  When $u \geq
d+1$ and $r-d+1 \leq v \leq r+d-1 $, a decoder error is possible,
and we determine the DEP based on the distance distribution of the
code with respect to $C$, denoted as $A_w(C)
\df |\{ D \in \mathcal{C}: \di(D, C) = w \}| = |\{ D \in
\mathcal{C}: \ds(D, C) = 2w \}|$.

\begin{proposition} \label{prop:PES}
Assuming a codeword $C$ of a CDC in $E_r(q,n)$ with minimum subspace distance $2d$ is sent over a symmetric operator channel and that the received subspace has dimension $v$ and is at subspace distance $u$ from $C$,
the DEP of a bounded subspace distance decoder with decoding radius $d-1$ is given by
\begin{equation} \label{eq:PES}
    \Ps(C,u,v, d-1) = \frac{1}{\Ns(r,v,u)}
    \sum_{w=d}^r A_w(C)
    \sum_{s=0}^{d-1} \Js(u,s,2w;r,r,v)
\end{equation}
when $\Ns(r,v,u) > 0$, $u \geq d+1$, and $|v-r| \leq d-1$; and $\Ps(C,u,v, d-1) = 0$ otherwise. Furthermore, if the CDC is the lifting
of a rank metric code, then
\begin{eqnarray}
	\nonumber
    &&\Ps(I({\bf C}),u,v, d-1)\\
    \label{eq:PES_lifting}
    &&\leq \frac{1}{\Ns(r,v,u)}
    \sum_{w=d}^r {r \brack w} \alpha(n-r,w-d+1)
    \sum_{s=0}^{d-1} \Js(u,s,2w;r,r,v)\\
    \label{eq:PES_lifting_bound}
    &&< \left\{ \begin{array}{ll}  L_q q^{-\frac{d-1+v-r}{2}\left(n-2r+\frac{d-1-v+r}{2}\right)} & \mbox{when } d-1+r-v
    \mbox{ is even},\\
    L_q q^{-\frac{d-1+v-r}{2}\left(n-2r+\frac{d-1+r-v}{2}\right) - \frac{1}{2}\left(n-d+1 + \frac{1}{2}\right)} & \mbox{when } d-1+r-v
    \mbox{ is odd}, \end{array} \right.
\end{eqnarray}
where $L_q = K_q^{-2} \sum_{i=0}^\infty q^{-\frac{3}{4}i^2}$.
\end{proposition}

The distinction on $d-1+r-v$ in the upper bound in (\ref{eq:PES_lifting_bound}) is explained as follows. For $d-1+r-v$ even, the largest subspace distance of a decodable subspace is exactly $d-1$; however, it is only $d-2$ when $d-1+r-v$ is odd. Therefore, for $v$ such that  $d-1+r-v$ is odd, the code uses a bounded distance decoder with true decoding radius $d-2$ instead of $d-1$, leading to a smaller DEP.

Note that the bound in (\ref{eq:PES_lifting_bound}) is very general, as it does not depend on the transmitted codeword or the distance of the received subspace to the sent codeword. We also remark that the exponent in the bound in (\ref{eq:PES_lifting_bound}) becomes zero when either $v = r-d+1$ or $r =\frac{n}{2}$ and $v=\frac{n}{2}+d-1$. The case where $v = r-d+1$ can be explained as follows. The decoding region around each codeword consists of decodable subspaces with dimension $v= r-d+1$, and hence is the intersection of the sphere of radius $d-1$ around a codeword and the Grassmannian $E_{r-d+1}(q,n)$. The volume of each decoding region is $\Ns(r,r-d+1,d-1) = {r \brack d-1}$, and for the lifting of an MRD code, the disjoint union of the decoding regions has $q^{(n-r)(r-d+1)} {r \brack d-1} \geq q^{(r-d+1)(n-r+d-1)}$ subspaces by (\ref{eq:Gaussian}). On the other hand, $|E_{r-d+1}(q,n)| < K_q^{-1} q^{(r-d+1)(n-r+d-1)}$ by (\ref{eq:Gaussian}), and hence the decoding regions form an asymptotically perfect packing of $E_{r-d+1}(q,n)$. Thus, the DEP, which is the ratio between the total number of decodable subspaces and $|E_{r-d+1}(q,n)|$, becomes zero. The case where $r = \frac{n}{2}$ and $v=\frac{n}{2}+d-1$ can be explained in a similar fashion.

We now show that liftings of MRD codes, referred to as KK codes,
have the highest DEP among all liftings up to a scalar in all
nontrivial cases. We denote the DEP of a BDD with decoding radius
$d-1$ for a KK code in $\mathrm{GF}(q)^{r \times (n-r)}$ with
minimum rank distance $d$ as $P_{\tiny \mbox{S,KK}}(u, v, d-1)$. Since
the distance distribution of a KK code is transparent to the
transmitted codeword, we have removed the dependence on the
transmitted codeword in the DEP for a KK code.
\begin{corollary} \label{cor:PES_KK}
Let $\mathcal{C}$ be any rank metric code in $\mathrm{GF}(q)^{r \times
(n-r)}$ ($r \leq n-r$) with minimum rank distance $d$ and let ${\bf
C} \in \mathcal{C}$. Then if $q>2$, $r<n-r$, or $d \neq n-r-1$,
$\Ps(I({\bf C}), u, v, d-1) < H_q P_{\tiny \mbox{S,KK}}(u, v, d-1)$, where
$H_2 = 3.5$ and $H_q = \frac{q-1}{q-2}$ for $q > 2$.
\end{corollary}
The proof of Corollary~\ref{cor:PES_KK} is similar to that of
Corollary~\ref{cor:PER_symmetric_MRD} and is hence omitted. Again
note that Corollary~\ref{cor:PES_KK} does not hold only for a trivial case.

If the probability that the received subspace is at subspace
distance $u$ from the sent codeword decreases rapidly with $u$, then
the overall DEP is dominated by $\Ps(C, d+i, v, d-1)$, where $i=1$
when $d-1+v-r$ is even and $i=2$ otherwise. Note that $u=d+i$ is the
smallest subspace distance that leads to a decoding error and
ensures $u+v-r$ is even and hence $\Ns(r,v,u) > 0$.
Proposition~\ref{prop:PES_d-t} below determines this value, and
shows that it asymptotically reaches the upper bound
in~(\ref{eq:PES_lifting_bound}) for KK codes.

\begin{proposition} \label{prop:PES_d-t}
The DEP of a bounded subspace distance decoder with decoding radius $d-1$ for a CDC in $E_r(q,n)$ with minimum subspace distance $2d$ over a symmetric operator
channel, provided that the received subspace is at subspace distance
$d+i$ ($i=1$ when $d-1+v-r$ is even and $i=2$ otherwise) from the
sent codeword, is given by
\begin{equation} \label{eq:PES_d-t}
    \Ps(C, d+i, v, d-1) = q^{-(d-\tau)(\tau+i)}
    \frac{ {d \brack \tau} {d \brack \tau+i} }{ {r \brack d-\tau} {n-r \brack \tau+i} }
    A_d(C),
\end{equation}
where $\tau = \frac{d-i+v-r}{2}$. In particular, the DEP for a KK code satisfies $P_{\tiny
\mbox{S,KK}}(d+1, v, d) > K_q^2 q^{-\frac{d-1+v-r}{2}\left(n-2r+\frac{d-1-v+r}{2}\right)}$ when
$d-1+v-r$ is even and $P_{\tiny \mbox{S,KK}}(d+2, v, d) > K_q^2
q^{-\frac{d-1+v-r}{2}\left(n-2r+\frac{d-1-v+r}{2}\right) - \frac{1}{2}\left(n-d+1 + \frac{1}{2}\right)}$ when $d-1+v-r$ is odd.
\end{proposition}

The proof of Proposition~\ref{prop:PES_d-t} is similar to that of
Proposition~\ref{prop:PER_d-t} and is hence omitted.
Proposition~\ref{prop:PES_d-t} and (\ref{eq:PES_lifting_bound})
indicate that the upper bounds in (\ref{eq:PES_lifting_bound}) for a
KK code are tight, since the lower bounds in
Proposition~\ref{prop:PES_d-t} and the upper bounds in
(\ref{eq:PES_lifting_bound}) differ by a scalar only.

\subsection{DEP of Bounded Injection Distance Decoders for CDCs}
\label{sec:PEM}

We now study the DEP of a bounded injection decoder with decoding radius $t = \left\lfloor \frac{d-1}{2} \right\rfloor$ for a CDC with minimum injection distance $d$ (and hence, minimum subspace distance $2d$) over a symmetric operator channel. Recall that if the received subspace has dimension $v$ and is at injection distance $\mu$ from the transmitted codeword, then it is at subspace distance $u = 2\mu - |v-r|$. Therefore, the subspaces decodable by the bounded injection distance decoder are at subspace distance no more than $d-1$ from a codeword and can all be decoded by the bounded subspace distance decoder of the same code. In other words, a bounded injection distance decoder only decodes a fraction of subspaces decodable by a bounded subspace distance decoder. We refine this statement and express the DEP of the bounded injection decoder for a code in terms of the DEP for the bounded subspace distance decoder of the same code in Proposition \ref{prop:Pi=Ps} below.

\begin{proposition} \label{prop:Pi=Ps}
A BDD with injection decoding radius $t$ corrects the same subspaces as a BDD with subspace decoding radius $2t-|v-r|$. Therefore, $\Pi(C,\mu,v,t) = \Ps(C,2\mu-|v-r|,v,2t-|v-r|)$.
\end{proposition}

\begin{proof}
A subspace of dimension $v$ is at injection distance $\mu$ from the sent codeword and at distance no more than $t$ than another codeword if and only if it is at subspace distance $u$ from the sent codeword and at subspace distance no more than $2t-|v-r|$ than another codeword. Therefore, a subspace is decodable by a BDD with injection radius $t$ if and only if it is decodable by a BDD with subspace radius $2t-|v-r|$.
\end{proof}



We hence apply the bounds on the DEP of the bounded subspace distance decoder to derive bounds on the DEP of the bounded injection distance decoder of the same code.

\begin{proposition} \label{prop:PEM}
Assuming a codeword $C$ of a CDC in $E_r(q,n)$ with minimum injection distance $d$ is sent over a symmetric operator channel and that the received subspace has dimension $v$ and is at injection distance $\mu$ from $C$,
the DEP of a bounded injection distance decoder with decoding radius $t$ is given by
\begin{equation} \label{eq:PEM}
    \Pi(C, \mu, v, t) = \frac{1}{\Ns(r,v,2\mu-|v-r|)} \sum_{w=d}^r A_w(C) \sum_{s=0}^{2t-|v-r|} \Js(2\mu-|v-r|,s,2w;r,r,v)
\end{equation}
for $\mu \geq d-t+ |v-r|$ and $|v-r| \leq t$, and $\Pi(C, \mu, v, t) = 0$ otherwise. Furthermore, if the CDC is the lifting of a rank metric code, then
\begin{equation}
    \label{eq:PEM_lifting_bound}
    \Pi(I({\bf C}), \mu, v, t) < \left\{ \begin{array}{ll}  L_q q^{-\left(t + \frac{v-r}{2}\right)\left(n-2r+t - \frac{v-r}{2}\right) - \frac{|v-r|}{2}\left(n-2t + \frac{|v-r|}{2} \right)} & \mbox{ when } d=2t+1,\\
    L_q q^{-\left(t + \frac{v-r}{2}\right)\left(n-2r+t - \frac{v-r}{2}\right) - \frac{|v-r|}{2}\left(n-2t + \frac{|v-r|}{2} \right) - n+r} & \mbox{ when }
    d=2t+2,
    \end{array} \right.
\end{equation}
where $L_q = K_q^{-2} \sum_{i=0}^\infty q^{-\frac{3}{4}i^2}$.
\end{proposition}

%

For the injection distance, there is no distinction on $v$ like in (\ref{eq:PES_lifting_bound}) for the subspace metric, because the true subspace distance decoding radius is always given by $2t-|v-r|$. Proposition~\ref{prop:PEM} indicates that at least $d-t$ errors and at least $d-t$ erasures both have to occur for the bounded injection distance decoder to decode erroneously. Therefore, the bounded injection distance decoder is more robust to errors. We finally remark that analogues of the other results derived for the subspace distance decoder can also be derived for the injection distance decoder, however we shall omit them for the sakes of clarity and conciseness.

\begin{figure}
\begin{center}
\includegraphics[scale=0.75]{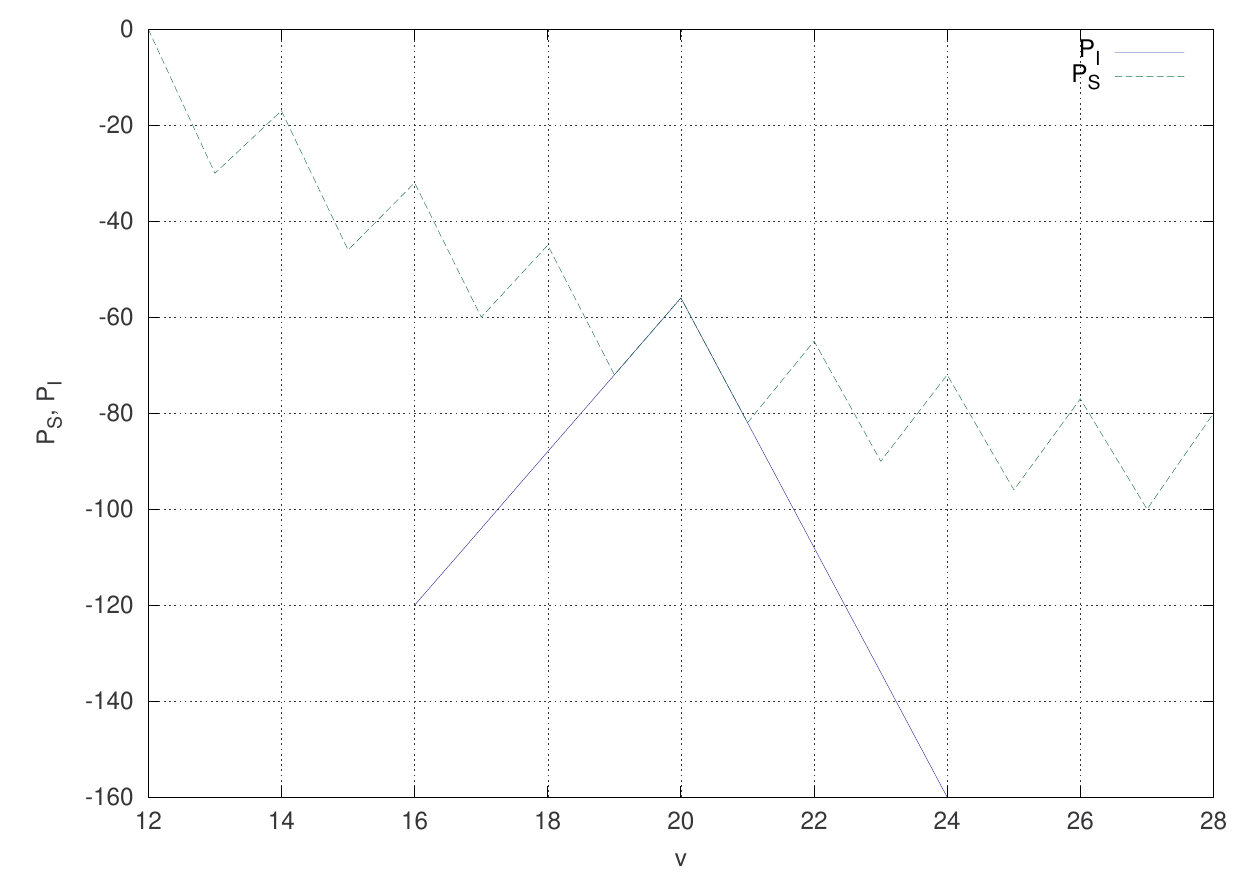}
\caption{DEP of BDDs with the subspace distance and the injection distance for $n=50$, $r=20$, $d=9$, $t=4$ ($\log_q$ scale).} \label{fig:pips}
\end{center}
\end{figure}

\begin{figure}[htb]
\begin{center}
\mbox{\subfigure[in projective space with subspace metric] {\scalebox{0.85}
{\includegraphics{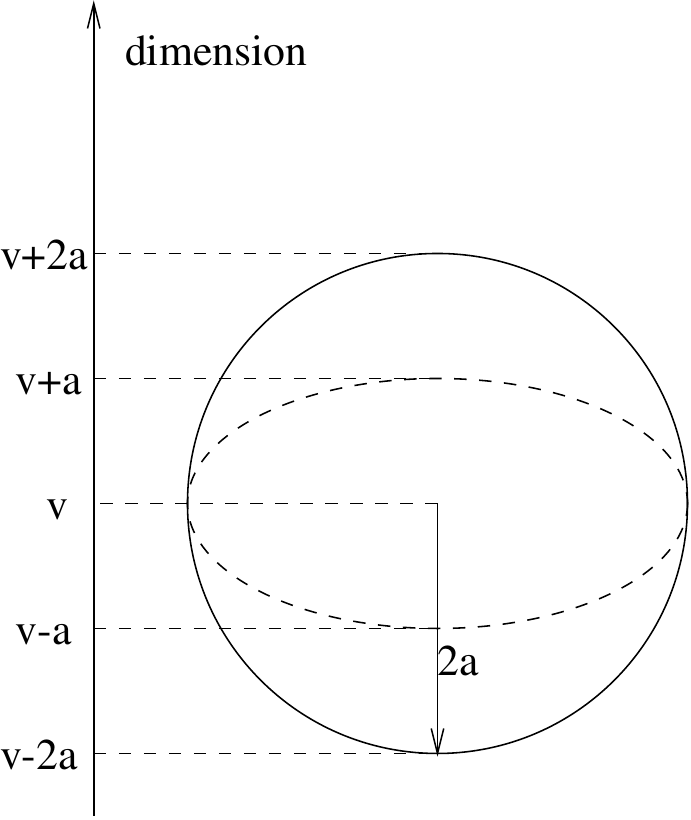}}}
\quad\quad \subfigure[in projective space with injection metric]
{\scalebox{1.00}
{\includegraphics{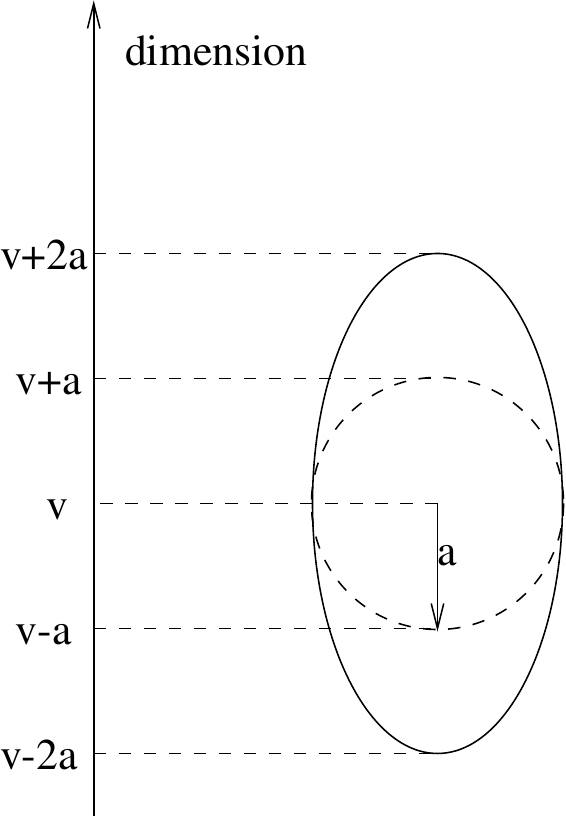}}}}
\caption{Comparison of the ball with subspace radius $2a$ (solid line) and the ball with injection radius $a$ (dashed line), both centered at a subspace with dimension $v$ (we assume $0\leq v-2a\leq v+2a \leq n$)\label{fig:relation}}
\end{center}
\end{figure}

The DEPs of the bounded subspace and injection decoders for the same code are compared in Figure \ref{fig:pips}. More precisely, we consider a CDC in $E_{20}(q,50)$ with minimum injection distance $d=9$ and error correction capability $t = 4$. the exponents in the bounds in (\ref{eq:PES_lifting_bound}) and (\ref{eq:PEM_lifting_bound}) are shown for $r-d+1 \leq v \leq r+d-1$ and $r-t \leq v \leq r+t$, respectively. We clearly see that both decoders have the same radius for $r-1 \leq v \leq r+1$, while their performances diverge for larger values of $|v-r|$. This illustrates the results in Proposition \ref{prop:Pi=Ps}, where it is shown that the bounded injection decoder can be viewed as a bounded subspace distance decoder whose decoding radius decreases with $|v-r|$. These relations are further illustrated in Figure \ref{fig:relation}, where we schematically compare the volumes of decoding spheres around codewords for both metrics.

\section{Acknowledgments}

We would like to thank the Associate Editor Dr. Mario Blaum and the anonymous reviewers for their valuable comments, some of which simplified the proofs of our results.

\appendix

\subsection{Proof of Proposition \ref{prop:PER_row}}
\label{app:prop:PER_row}

In order to prove Proposition~\ref{prop:PER_row}, we first need a technical lemma.

\begin{lemma}\label{lemma:sum_g}
For all $U \in E_u(q,n)$, $\sum_{W \in E_w(q,n)} \gr(U, s, W) =
\frac{{n \brack w}}{{n \brack u}} \Jr(u,s,w)$.
\end{lemma}

\begin{proof}
For any $A\in E_a(q,n)$, the number of matrices in
$\mathrm{GF}(q)^{m \times n}$ with row space equal to $A$ is given
by $\alpha(m,a)$. Hence by counting the number of pairs of matrices
$({\bf R}, {\bf W})$, where ${\bf R}, {\bf W}\in \mathrm{GF}(q)^{m \times n}$,  $R({\bf R}) = U$, $R({\bf W}) = W$, and
$\dr({\bf R}, {\bf W}) = s$, in two ways, we have $\alpha(m,u) \gr(W,
s, U) = \alpha(m,w) \gr(U, s, W)$. Hence $\sum_{W \in E_w(q,n)} \gr(U,
s, W) = \frac{\alpha(m,u)}{\alpha(m,w)} \sum_{W \in E_w(q,n)} \gr(W,
s, U) = \frac{\alpha(m,u)}{\alpha(m,w)} \Jr(w,s,u)$. Using
(\ref{eq:N_J}), we obtain $\sum_{W \in E_w(q,n)} \gr(U, s, W) =
\frac{{n \brack w}}{{n \brack u}} \Jr(u,s,w).$
\end{proof}

We now give the proof of Proposition~\ref{prop:PER_row}, whose outline is as follows. We first prove (\ref{eq:PER_row_exact}) by counting the number of decodable matrices. Then, deriving an upper bound on the distribution of rank metric codes leads to (\ref{eq:PER_row_bound}). Finally, we use geometric properties of the rank metric and Lemma \ref{lemma:sum_g} to obtain the general upper bound in (\ref{eq:PER_row_asymptotic}).

\begin{proof}
Let $\mathcal{C}$ be a rank metric code in $\mathrm{GF}(q)^{m \times
n}$ with minimum rank distance $d$. We have $\Pr({\bf C}, U) =
\frac{1}{\alpha(m,u)} \sum_{s=0}^t \delta(U, s, {\bf C})$, where
$\delta(U, s, {\bf C})$ is the number of matrices ${\bf X}$ at
distance $s$ from the code, and such that $R({\bf X} - {\bf C}) =
U$. Hence $\delta(U, s, {\bf C}) = \sum_{W: \dim(W) \geq d} A_W({\bf
C}) \gr(U, s, W)$ and we obtain (\ref{eq:PER_row_exact}).

We now give an upper bound on $A_W({\bf C})$. We can express any
matrix ${\bf E} \in \mathrm{GF}(q)^{m \times n}$ with row space $W
\in E_w(q,n)$ as ${\bf E} = {\bf F} {\bf W}$, where ${\bf F} \in
\mathrm{GF}(q)^{m \times w}$ has rank $w$ and ${\bf W} \in \mathrm{GF}(q)^{w
\times n}$ is a fixed matrix with row space $W$. Let $\mathcal{C}_W
= \{ {\bf F}: {\bf F}{\bf W} + {\bf C} \in \mathcal{C}\}$, then
$|\mathcal{C}_W| = A_W({\bf C})$ and $\mathcal{C}_W$ forms a
constant-rank code in $\mathrm{GF}(q)^{m \times w}$ with rank $w$
and minimum distance $d$. Therefore, $A_W({\bf C}) \leq
\Ar(q,m,w,d,w) \leq \alpha(m,w-d+1)$ by (\ref{eq:bound_Ar}). The DEP
hence satisfies
\begin{eqnarray}
    \nonumber
    \Pr({\bf C}, U, t) 
    &\leq& \frac{1}{\alpha(m,u)} \sum_{s=0}^t \sum_{w=d}^n
    \alpha(m,w-d+1) \sum_{W \in E_w(q,n)}
    \gr(U,s,W)\\
    \label{eq:P3}
    &=& \frac{1}{\Nr(u)}  \sum_{w=d}^n {n \brack w}
    \alpha(m,w-d+1) \sum_{s=0}^t \Jr(u,s,w),
\end{eqnarray}
where (\ref{eq:P3}) follows Lemma~\ref{lemma:sum_g}.

When $C$ is an MRD code with minimum rank distance $d$, it can be
shown that $A_W({\bf C}) = M(q,m,w,d,w) = {n \brack w}^{-1}
M(q,m,n,d,w)$ \cite[Lemma 3]{gabidulin_pit0185},
which leads to (\ref{eq:PER_row_MRD}).

By definition of $\alpha(m,u)$,
\begin{equation}
    \label{eq:bound_alpha}
    {n \brack w} \alpha(m,w-d+1)
    = {n \brack w} \alpha(m,w)
    \frac{q^{-(d-1)(w-d+1)}} {\alpha(m-w+d-1,d-1)}
    < K_q^{-1} q^{-m(d-1)} \Nr(w),
\end{equation}
and hence
\begin{eqnarray}
    \nonumber
    \Pr({\bf C}, u, t) &<& \frac{1}{\Nr(u)} K_q^{-1} q^{-m(d-1)}
    \sum_{s=0}^t \sum_{w=d}^n \Nr(w) \Jr(u,s,w)\\
    \label{eq:P1}
    &=& K_q^{-1} q^{-m(d-1)} \sum_{s=0}^t \sum_{w=d}^n \Jr(w,s,u)\\
    \label{eq:P2}
    &\leq& K_q^{-1} q^{-m(d-1)} \Vr(t),
\end{eqnarray}
where (\ref{eq:P1}) and (\ref{eq:P2}) follow (\ref{eq:N_J}) and (\ref{eq:sum_J}), respectively. Using $\Vr(t) < K_q^{-1} q^{t(m+n-t)}$ \cite{gadouleau_it08_dep}, we obtain (\ref{eq:PER_row_asymptotic}).
\end{proof}

\subsection{Proof of Proposition \ref{prop:Js_triangle}} \label{app:prop:Js_triangle}

\begin{proof}
The outline of the proof is as follows. For $A \in E_a(q,n)$ and $B \in E_b(q,n)$ with $\ds(A,B) = u+s$, we want to count the number of $C \in E_c(q,n)$ satisfying
$\ds(A,C) = u$, and $\ds(B,C) = s$. We first prove that $C \cap A \cap B = A \cap B$ and $C = C \cap A + C \cap B$. We can then count the number of possible choices for $C \cap A$ and $C \cap B$, which yields the number of possible choices for $C$.

We have $\dim(C \cap A) = \frac{a+c-u}{2} = a'$, $\dim(C \cap B) = \frac{b+c-s}{2} = b'$, and $\dim(A \cap B) = \frac{a+b-u-s}{2} = a' + b' - c$. This implies that $\Js(u,s,u+s;a,b,c)=0$ when $u+s>a+b$, $u>a+c$, or $s>b+c$. Since
\begin{eqnarray}
	\nonumber
	\dim(C \cap A \cap B) &=& \dim(C \cap A) + \dim(C \cap B) - \dim(C \cap A + C \cap B)\\
	\nonumber
	&\geq& a' + b' - c = \dim(A \cap B),
\end{eqnarray}
we have $C \cap A \cap B = A \cap B$ and $\dim(C \cap A + C \cap B) = c$, hence $C = C \cap A + C \cap B$.

Since $C \cap A$ is a subspace of dimension $a'$ of $A$ which contains $A \cap B$, there are $\Ns(q,a,a'+b'-c, a', c-b')$ choices for $C \cap A$. Note that $\Ns(q,a,a'+b'-c, a', c-b')=0$ when $s>2a+b-c$. Similarly, there are $\Ns(q,b,a'+b'-c, b', c-a')$ choices for $C \cap B$. Note that $\Ns(q,b,a'+b'-c, b', c-a')=0$ when $u>2b+a-c$. Thus, there are $\Ns(q,a,a'+b'-c, a', c-b') \Ns(q,b,a'+b'-c, b', c-a')={a-a'-b'+c \brack a'} {b-a'-b'+c \brack b'}$ choices for $C$.
\end{proof}

\subsection{Proof of Proposition \ref{prop:PES}} \label{app:prop:PES}

We first prove a technical lemma.

\begin{lemma} \label{lemma:sum_Ns}
For all $r$, $s$, and $t \leq \min \left\{r+s, \left\lfloor \frac{n}{2} \right\rfloor \right\}$, $\sum_{d=0}^t \Ns(r,s,d) < L_q q^{f(r,s,t)}$, where $4f(r,s,t) = t(2n-t) - (r-s)(2n-r-3s)$ and $L_q =  K_q^{-2} \sum_{i=0}^\infty q^{-\frac{3}{4}i^2}$.
\end{lemma}

\begin{proof}
By definition of $\Ns(r,s,d)$ and (\ref{eq:Gaussian}), we have $\Ns(r,s,d) < K_q^{-2} q^{f(r,s,d)}$, and hence $\sum_{d=0}^t \Ns(r,s,d) < K_q^{-2} q^{f(r,s,t)} \sum_{i=0}^t q^{-\frac{i}{4}(2n-2t+i)} \leq K_q^{-2} q^{f(r,s,t)} \sum_{i=0}^t q^{-\frac{3}{4}i^2} < L_q q^{f(r,s,t)}$.
\end{proof}

We now prove Proposition \ref{prop:PES}, whose outline is similar to that of Proposition \ref{prop:PER_row}. After proving the equality (\ref{eq:PES}), we give an upper bound on the distance distribution of liftings of rank metric codes to obtain (\ref{eq:PES_lifting}). We finally use geometric properties of the projective space and Lemma \ref{lemma:sum_Ns} to obtain the general upper bound in (\ref{eq:PES_lifting_bound}).

\begin{proof}
Given $u$ and $v$ such that $\Ns(r,v,u) > 0$, we have $\Ps(C, u, v, d-1) = \frac{D(C, u, v)}{\Ns(r,v,u)}$, where $D(C,
u, v)$ is the number of decodable subspaces with dimension $v$ and
at subspace distance $u$ from $C$. A subspace is decodable if it is within the decoding radius of any codeword. For a codeword $C'$ at subspace
distance $2w$ from $C$, there are exactly $\sum_{s=0}^{d-1}
\Js(u,s,2w; r,r,v)$ subspaces with dimension $v$, at distance $u$
from $C$, and at distance $\leq d-1$ from $C'$ by
Proposition~\ref{prop:Js}. Summing for all $C'$, we obtain (\ref{eq:PES}).

Let $I(\mathcal{C})$ be the lifting of a rank metric code
$\mathcal{C}$ in $\mathrm{GF}(q)^{r \times (n-r)}$. For ${\bf C} \in
\mathcal{C}$, $\{ I({\bf D} - {\bf C}): {\bf D} \in \mathcal{C},
\dr({\bf D}, {\bf C}) = w \}$ is the lifting of a constant-rank code in
$\mathrm{GF}(q)^{r \times (n-r)}$ with minimum rank distance at
least $d$ and rank $w$, and hence $A_w(I({\bf C})) \leq
\Ar(q,n-r,r,d,w) \leq {r \brack w} \alpha(n-r,w-d+1)$ by
(\ref{eq:bound_Ar}), which leads to (\ref{eq:PES_lifting}).

We now
prove (\ref{eq:PES_lifting_bound}) when $d-1+v-r$ is even, and the other
case is similar and its proof is omitted. We have $\alpha(n-r,w-d+1) \leq q^{-(n-r)(d-1)}
q^{w^2} {n-r \brack w}$ and $\Ns(r,r,2w) = q^{w^2} {r \brack w} {n-r
\brack w}$ and hence
\begin{equation}\label{eq:bound_Aw}
    A_w(I({\bf C})) \leq q^{-(n-r)(d-1)} \Ns(r,r,2w).
\end{equation}
We obtain
\begin{eqnarray}
    \label{eq:Ps1}
    \Ps(I({\bf C}), u, v, d-1) &=& \sum_{w=d}^r A_w(I({\bf C})) \sum_{s=0}^{d-1}
    \frac{\Js(2w,s,u;r,v,r)}{\Ns(r,r,2w)}\\
    \label{eq:Ps2}
    &\leq& q^{-(n-r)(d-1)} \sum_{w=d}^r \sum_{s=0}^{d-1} \Js(2w,s,u;r,v,r)\\
    \label{eq:Ps3}
    &\leq& q^{-(n-r)(d-1)} \sum_{s=0}^{d-1} \Ns(v,r,s),
\end{eqnarray}
where (\ref{eq:Ps1}), (\ref{eq:Ps2}), and (\ref{eq:Ps3}) follow
(\ref{eq:scaling_Js}), (\ref{eq:bound_Aw}), (\ref{eq:sum_Js}),
respectively. We obtain
(\ref{eq:PES_lifting_bound}) since $\sum_{s=0}^{d-1} \Ns(v,r,s) < L_q q^{\frac{1}{4}(d-1)(2n-d+1) - \frac{1}{4}(v-r)(2n-v-3r)}$ by Lemma \ref{lemma:sum_Ns}.
\end{proof}

\bibliographystyle{IEEEtran}
\bibliography{gpt}

\end{document}